\documentclass[twocolumn,showpacs,preprintnumbers,amsmath,amssymb]{revtex4}
\usepackage{graphicx}
\usepackage{dcolumn}
\usepackage{bm}
\def\bea {\begin{eqnarray}}
\def\eea {\end{eqnarray}}

\def\be {\begin{equation}}
\def\ee {\end{equation}}

\begin{document}

\title{Nuclear suppression at low energy heavy ion collisions}

\author{Santosh K Das, Jan-e Alam,  Payal Mohanty and Bikash Sinha}

\medskip

\affiliation{Variable Energy Cyclotron Centre, 1/AF, Bidhan Nagar , 
Kolkata - 700064}

\date{\today}
\begin{abstract}
The effects of non-zero baryonic chemical potential on the drag and diffusion
coefficients of heavy quarks propagating through a  baryon rich 
quark gluon plasma have been studied. The nuclear suppression factor, 
$R_{\mathrm AA}$ for non-photonic single electron spectra resulting from 
the semileptonic decays of hadrons containing heavy flavours have been 
evaluated for low  energy collisions. The role of non-zero baryonic chemical 
potential on $R_{\mathrm AA}$ has been highlighted.
\end{abstract}

\maketitle

\section{Introduction}
The nuclear collisions at low energy RHIC run~\cite{lowephenix,lowestar}
and GSI-FAIR~\cite{cbm}  is expected to
create a thermal medium with  large baryonic chemical potential ($\mu_B$)
and moderate temperature ($T$). 
The heavy flavours namely, charm
and bottom quarks may play a crucial role in understanding the
properties of such medium  because they do not constitute the bulk part   
of the system and their thermalization time scale
is larger than the light quarks and gluons and hence can retain the
interaction history very effectively.   
The perturbative QCD (pQCD) calculations indicate that the heavy quark ($Q$)
thermalization time, $\tau_i^Q$ is larger~\cite{moore,baier} than the 
light quarks 
and gluons thermalization scale $\tau_i$. 
Gluons may thermalized even before 
up and down quarks~\cite{japrl,shuryak}. In the present work we 
assume that the Quark Gluon Plasma (QGP)
is formed at time $\tau_i$. Therefore, the interaction
of the non-equilibrated heavy quarks with the equilibrated QGP for
the time interval $\tau_i<\tau<\tau_i^Q$ can be treated within the ambit 
of the Fokker-Planck (FP) equation~\cite{landau,balescu}
{\it i.e.} the heavy quark can be thought of
executing Brownian motion~\cite{moore,japrl,sc,svetitsky,rapp,turbide,
bjoraker,npa1997,munshi,rma} 
in the heat bath of QGP during the said interval of time.

As the relaxation time
for heavy quarks of mass $M$  at a temperature
$T$ are larger than the corresponding quantities
for light partons by a factor of $M/T(>1)$~\cite{moore} 
{\it i.e.} the
light quarks and the gluons get thermalized faster than the heavy 
quarks, the propagation of  heavy quarks through 
QGP (mainly contains light quarks and gluons) therefore, 
may be treated as the 
interactions between equilibrium and non-equilibrium degrees
of freedom. The FP equation provide an appropriate frame work for such
processes. In case of low energy collisions the radiative energy loss of 
of heavy quarks will be much smaller than the loss 
due to elastic processes. Moreover,
the thermal production of charm and bottom quarks can be ignored 
for the range of temperature and
baryonic chemical potential under study. 
Therefore, the FP equation is better 
applicable in the present situation.

The paper is organized as follows. In the next section a brief
description of Fokker Planck equation and 
the $T$ and $\mu_q$ dependence of the drag and diffusion 
co-efficients are outlined, the non-photonic electron spectra 
is discussed in section III, the initial conditions and the space 
time evolution have been discussed in section IV, section V is devoted to 
the discussions on nuclear suppression and
finally section VI contains the summary and conclusions. 

\section{The Fokker Planck Equation}
The Boltzmann transport equation describing a non-equilibrium 
statistical system reads:
\be
\left[\frac{\partial}{\partial t} 
+ \frac{\bf p}{E}.\bf{\nabla_x} 
+ {\bf F}.\bf{\nabla_p}\right]f(x,p,t)=
\left[\frac{\partial f}{\partial t}\right]_{col}
\ee
where $p$ and $E$ denote momentum and energy, ${\bf{\nabla_x}}$
(${\bf{\nabla_p}}$) are spatial (momentum space) gradient and $f(x,p,t)$
is the phase space distribution (in the present case $f$ stands for
heavy quark distribution).
The assumption of uniformity in the plasma and absence of any external force
leads to
\be 
\frac{\partial f}{\partial t}=
\left[\frac{\partial f}{\partial t}\right]_{\mathrm col}
\ee
The collision term on the right hand side of the above equation can be 
approximated as (see ~\cite{svetitsky,npa1997} for details): 
\be
\left[\frac{\partial f}{\partial t}\right]_{col} = 
\frac{\partial}{\partial p_i} \left[ A_i(p)f + 
\frac{\partial}{\partial p_i} \lbrack B_{ij}(p) f \rbrack\right] 
\label{expeq}
\ee
where we have defined the kernels 
\begin{eqnarray}
&& A_i = \int d^3 k \omega (p,k) k_i \nonumber\\
&&B_{ij} = \int d^3 k \omega (p,k) k_ik_j.
\end{eqnarray}
for $\mid\bf{p}\mid\rightarrow 0$,  $A_i\rightarrow \gamma p_i$ 
and $B_{ij}\rightarrow D\delta_{ij}$ where $\gamma$ and $D$ stand for
drag and diffusion co-efficients respectively.
The function $\omega(p,k)$ is given by
\be
\omega(p,k)=g\int\frac{d^3q}{(2\pi)^3}f^\prime(q)v\sigma_{p,q\rightarrow p-k,q+k}
\ee
where $f^\prime$ is the phase space distribution, in the present
case it stands for light quarks and gluons,
$v$ is the relative velocity between the two collision partners,
$\sigma$ denotes the cross section and $g$ is the statistical
degeneracy. The co-efficients in the first two terms of the expansion
in Eq.~\ref{expeq} are comparable in magnitude because the averaging
of $k_i$ involves greater cancellation than the averaging of the
quadratic term $k_ik_j$. The higher power of $k_i$'s are smaller~\cite{landau}.

With these approximations the Boltzmann equation reduces to a non-linear
integro-differential equation known as Landau 
kinetic equation:
\be
\frac{\partial f}{\partial t} = 
\frac{\partial}{\partial p_i} \left[ A_i(p)f + 
\frac{\partial}{\partial p_i} \lbrack B_{ij}(p) f\rbrack \right] 
\label{landaueq}
\ee
The nonlinearity is caused due to the
appearance of $f^\prime$ in $A_i$ and $B_{ij}$ through $w(p,k)$.
It arises from the simple fact that we are studying
a collision process which involves two particles - it should,
therefore, depend on the states of the two participating particles in
the collision process and hence on the product of the two distribution 
functions.
Considerable simplicity may be achieved by replacing the distribution
functions of one of the collision partners by their 
equilibrium Fermi-Dirac or Bose-Einstein distributions
(depending on the statistical nature)
in the expressions of $A_i$ and $B_{ij}$. Then Eq.~\ref{landaueq} 
reduces to a linear
partial differential equation - usually referred to as the Fokker-Planck
equation\cite{balescu} describing the interaction of a particle which
is out of thermal equilibrium with the particles in a thermal bath
of light quarks, anti-quarks and gluons.
The quantities $A_i$ and $B_{ij}$ are related to the usual 
drag and diffusion coefficients and we denote them by $\gamma_i$ and
$D_{ij}$ respectively ({\it i.e.} these quantities can be obtained
from the expressions for $A_i$ and $B_{ij}$ by replacing the distribution
functions by their thermal counterparts):

The evolution of the heavy quark distribution ($f$) is governed by the 
FP equation~\cite{svetitsky}
\be
\frac{\partial f}{\partial t} =
\frac{\partial}{\partial p_i} \left[ \gamma_i(p)f +
\frac{\partial}{\partial p_i} \lbrack D_{ij}(p) f \rbrack \right]
\label{FPeq}
\ee
where $\gamma_i$ and  $D_{ij}$ are the drag and diffusion coefficients. 
The elastic collisions of heavy quarks ($Q$ stands for charm and
bottom) with thermal light quarks ($q$), anti-quarks ($\bar{q}$)
and gluons ($g$) {\it i.e.} 
$Qq \rightarrow Qq$, $Q\bar{q} \rightarrow Q\bar{q}$ 
and $Qg \rightarrow Qg$ 
have been used to evaluate
the drag and diffusion coefficients as indicated in~\cite{svetitsky,Das}. 
The thermal distribution of 
the quarks and the anti-quarks  are responsible   
for the $T$ and $\mu_q$ dependence of the drag
and diffusion co-efficients. 
The gluon distribution introduces the $T$ dependence 
to these quantities.
chemical potential dependence of the drag and diffusion 
co-efficients originate from the thermal phase space of 
quarks and anti-quarks.

\subsection{The drag and diffusion co-efficients}
At low $\sqrt{s_{\mathrm NN}}$ the net baryon
density at mid-rapidity is non-zero and its value
could be high depending on the value of   
 $\sqrt{s_{\mathrm NN}}$. Therefore, we need to solve
the FP equation for non-zero $\mu_B$. The drag and diffusion
coefficients are functions of both the thermodynamical variables:
$\mu_B$ and $T$. 

The energy dependence of the chemical potential 
has been  obtained from the parametrization of the  experimental data on
hadronic ratios as~\cite{ristea} (see also~\cite{andronic}),
\be
\mu_B(s_{\mathrm NN})=a(1+\sqrt{s_{\mathrm NN}}/b)^{-1}
\label{mub}
\ee
where $a = 0.967\pm 0.032$ GeV and $b=6.138 \pm 0.399$ GeV.
The parametrization in Eq.~\ref{mub} gives the values
of $\mu_B$ at the freeze-out. The corresponding values
at the initial condition are obtained from the 
baryon number conservation equation.  
The initial baryonic chemical potential 
carried by the  quarks $\mu_q (=\mu_B/3)$ are shown in table 1 for 
various $\sqrt{s_{\mathrm NN}}$ under consideration.

\begin{figure}[h]
\begin{center}
\includegraphics[scale=0.43]{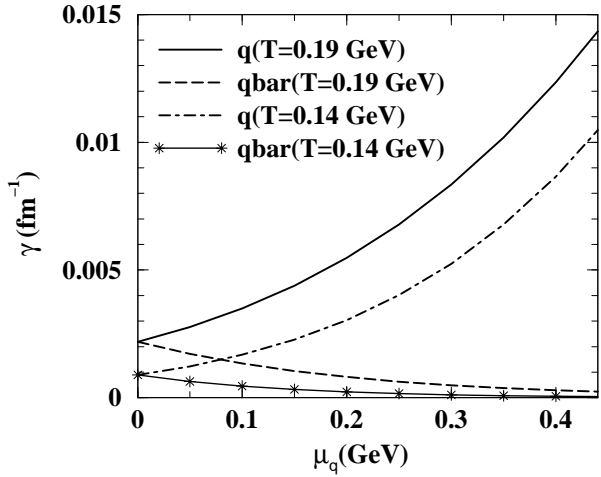}
\caption{Variation of the drag coefficient of charm quark due to its
interactions with light quarks and anti-quarks as a 
function of  $\mu_q$ for different temperatures.
}
\label{fig1}
\end{center}
\end{figure}

In the present work we take $\alpha_s=0.3$ because
the dependence of the strong coupling on the
temperature and baryonic chemical potential is not accurately known yet.
The sensitivity of collisional
energy loss on the running $\alpha_s$ is studied 
in Ref.~\cite{gossiaux} in detail.
The variation of the drag coefficients of charm quarks 
(due to its interactions with quarks and anti-quarks) 
on the baryonic chemical 
potential  for different $T$ are displayed in Fig.~\ref{fig1}.
The drag co-efficient for the process : $Qg\rightarrow Qg$
is  $\sim 8.42\times 10^{-3}$ fm$^{-1}$ 
($1.86\times 10^{-2}$ fm$^{-1}$) for  $T=140$ MeV (190 MeV)
(not displayed in Fig.~\ref{fig1}).
The $T$ and $\mu_q$ dependence of the drag and diffusion co-efficients
may be understood as follows.
The drag may be defined as the thermal average of the
square of the invariant transition amplitude weighted by
the momentum transfer for the reactions $qQ\,\rightarrow\, qQ$,  
$Q\bar{q}\,\rightarrow\, Q\bar{q}$ and
$gQ\,\rightarrow\,gQ$. As the temperature of the thermal bath
increases the light quarks ($q$) and the gluons move faster
and gain the ability to transfer larger momentum during their
interaction with the heavy quarks - resulting in
the increase of the drag of the heavy quarks propagating through
the partonic medium. Since the average momentum of the 
quarks increases with $\mu_q$,
similar behaviour is expected
in the variation of drag with  
baryonic chemical potential. This trend is clearly observed 
in the results displayed in Fig.~\ref{fig1} for 
charm quark.  The drag due to the process $Qq\rightarrow Qq$ is 
larger than the  $Q\bar{q}\rightarrow Q\bar{q}$ interaction
because for non-zero chemical potential, the $Q$ propagating
through the medium encounters more $q$ than $\bar{q}$ at a given $\mu_q$. 
For vanishing
chemical potential the contributions from quarks and anti-quarks 
are same.  

\begin{figure}[h]
\begin{center}
\includegraphics[scale=0.43]{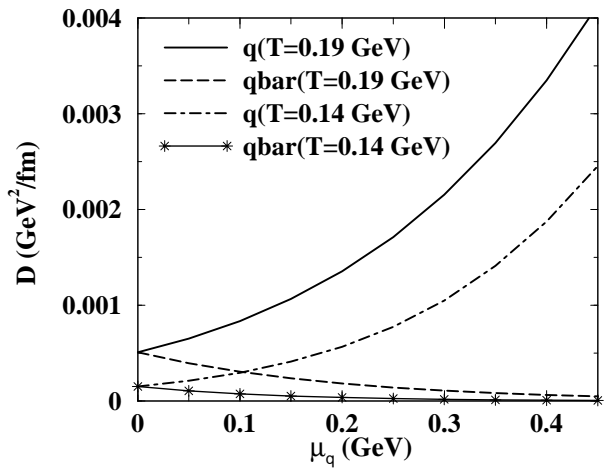}
\caption{Variation of the diffusion coefficient of charm quark due to its
interactions with light quarks and anti-quarks as a 
function of  $\mu_q$ for different temperatures.
}
\label{fig2}
\end{center}
\end{figure}

\begin{figure}[h]
\begin{center}
\includegraphics[scale=0.43]{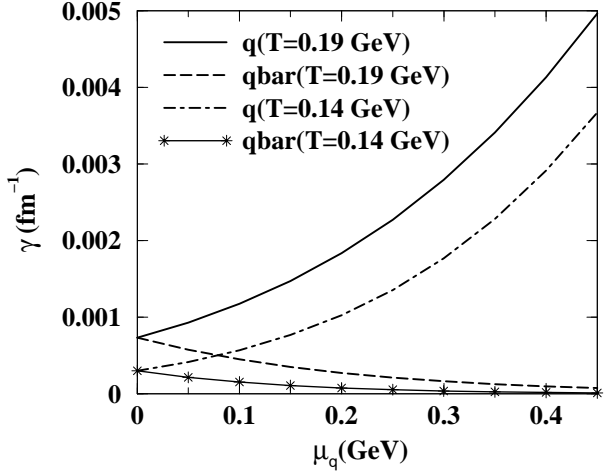}
\caption{Same as Fig.~\protect{\ref{fig1}} for bottom quark.
}
\label{fig3}
\end{center}
\end{figure}

\begin{figure}[h]
\begin{center}
\includegraphics[scale=0.43]{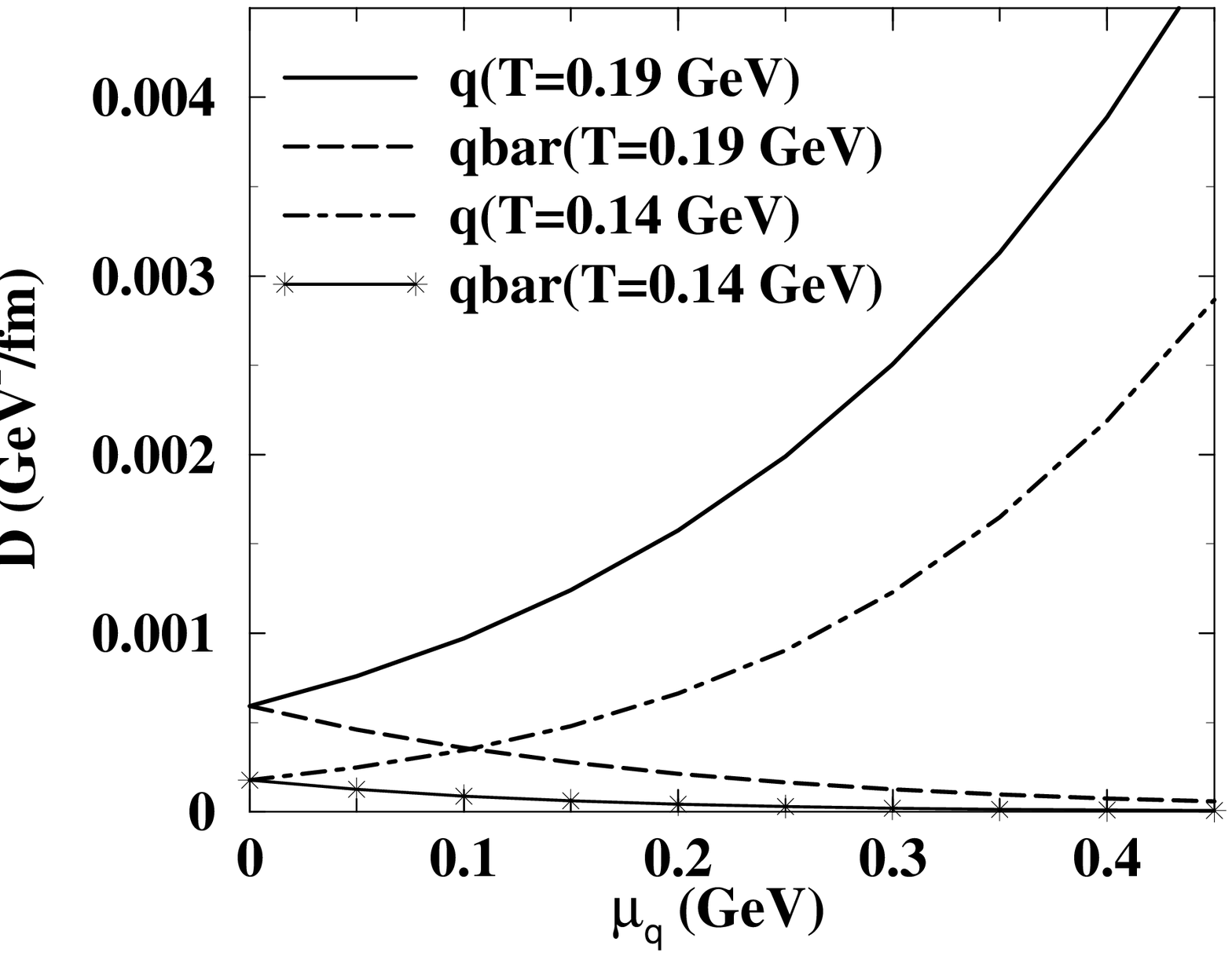}
\caption{Same as Fig.~\protect{\ref{fig2}} for bottom quark.
}
\label{fig4}
\end{center}
\end{figure}

In the same way it may
be argued that the diffusion coefficient involves the square of
the momentum transfer - which should also increase with
$T$ and $\mu_q$ as observed in Fig.~\ref{fig2}. The diffusion
co-efficient for charm quarks due to its interaction
with gluons is given by
 $\sim 1.42\times 10^{-3}$ GeV$^2$/fm 
($4.31\times 10^{-3}$ GeV$^2$/fm) for $T=140$ MeV (190 MeV).
It may be mentioned here that the drag increases with 
$T$ when the system
behave like a gas. In case of liquid the
drag may decrease with temperature (except very few cases) - 
because a substantial part of the thermal
energy goes in making the  attraction between the interacting
particles weaker - allowing them to move more freely  and hence
making the drag force lesser. 
The drag co-efficient of the partonic system with 
non-perturbative effects   
may decrease with temperature as shown in Ref.~\cite{rapphees} - because
in this case the system interacts strongly  more like a liquid.
The heavy quark momentum diffusion co-efficient has been
computed~\cite{caron-huot} at next to leading order within the
ambit of hard thermal loop approximations. For $T\sim 400$ MeV
the momentum averaged pQCD  value (for $\mu_q=0$) 
of the diffusion co-efficient obtained in the present
work is comparable to the value
obtained in~\cite{caron-huot} in the leading order approximation
for the same set of inputs ({\it e.g.} strong coupling constant,
number of flavours etc).
The drag and diffusion coefficients for bottom quarks are
displayed in Figs.~\ref{fig3} and \ref{fig4} respectively,
showing qualitatively similar behaviour as charm quarks.
The drag co-efficients for bottom quarks due to the process 
$Qg\rightarrow Qg$ 
is given by $\sim 3.15\times 10^{-3}$ fm$^{-1}$ and   
$6.93\times 10^{-3}$ fm$^{-1}$ at $T$ = 140 MeV and 190 MeV respectively.
The corresponding diffusion coefficients 
are $\sim 1.79\times 10^{-3}$ GeV$^2$/fm  and
$5.38\times 10^{-3}$ GeV$^2$/fm at T=140 MeV and 190 MeV respectively.

\section{The non-photonic electron spectra}
After obtaining the drag and diffusion coefficients 
we need the initial heavy quark momentum distributions 
for solving the FP equation.  
For low collision energy rigorous QCD based calculations for 
heavy flavour production is not available 
(for higher $\sqrt{s_{NN}}=200$ GeV see Ref.~\cite{matteo} for rigorous 
QCD calculations). In the present work 
this is obtained from pQCD calculation~\cite{pqcd,combridge}
for the processes: $gg\rightarrow Q\bar{Q}$
and $q\bar{q}\rightarrow Q\bar{Q}$.  Here we intend to deal
with the nuclear suppression factor, $R_{\mathrm AA}$, which
involves the ratio of the momentum distribution functions. Therefore,
the final results may not be too sensitive to the initial distributions 
because of some cancellations that may take place in the ratio. 

With the initial condition mentioned above 
the FP equation has been solved for the heavy quarks. We
convolute the solution with the fragmentation functions of the
heavy quarks to obtain the $p_T$ distribution of the
heavy ($B$ and $D$) mesons ($dN^{D,B}/q_Tdq_T$). For heavy quark 
fragmentation 
we use Peterson function~\cite{peterson} given by:
\be
f(z) \propto 
\frac{1}{\lbrack z \lbrack z- \frac{1}{z}- \frac{\epsilon_c}{1-z} \rbrack^2 \rbrack}
\ee
for charm quark $\epsilon_c=0.05$. For bottom quark 
$\epsilon_b=(M_c/M_b)^2\epsilon_c$ where $M_c$ ($M_b$) is the charm
(bottom) quark mass.

The non-photonic single electron spectra originate from the 
decays of heavy flavour mesons - {\it e.g.} $D\rightarrow Xe\nu$ 
at mid-rapidity ($y=0$)
can be obtained as follows~\cite{gronau,ali,akc}:
\be
\frac{dN^e}{p_Tdp_T}=\int dq_T \frac{dN^D}{q_Tdq_T} F(p_T,q_T)
\ee
where
\be
F(p_T,q_T)=\omega\int \frac{d(\bf{p}_T.\bf{q}_T)}{2p_T\bf{p}_T.\bf{q}_T}g(\bf{p}_T.\bf{q}_T/M)
\ee
where $M$ is the mass of the heavy mesons ($D$ or $B$),
$\omega=96(1-8m^2+8m^6-m^8-12m^4lnm^2)^{-1}M^{-6}$ ($m=M_X/M$) and $g(E_e)$ is given by
\be
g(E_e)=\frac{E_e^2(M^2-M_X^2-2ME_e)^2}{(M-2E_e)}
\ee
related to the  rest frame spectrum for the decay $D\rightarrow X e \nu$
through the following relation~\cite{gronau}
\be
\frac{1}{\Gamma}\frac{d\Gamma}{dE_e}=\omega g(E_e).
\ee

We evaluate the electron spectra from the decays of heavy mesons
originating from the fragmentation of the heavy quarks propagating
through the QGP medium formed in heavy ion collisions.
In the same way the electron spectrum from the p-p collisions
can be obtained from the charm and bottom quark distribution
which goes as initial conditions to the solution of FP equation.
The ratio of these two quantities gives the nuclear suppression,
$R_{AA}$ as :
\be
R_{AA}(p_T)=\frac{\frac{dN^e}{d^2p_Tdy}^{\mathrm Au+Au}}
{N_{\mathrm coll}\times\frac{dN^e}{d^2p_Tdy}^{\mathrm p+p}}
\label{raa}
\ee
called the nuclear suppression factor,
will be unity in the absence of any medium. In the above equation 
$N_{\mathrm {coll}}$ denotes the number of nucleon nucleon collisions
in Au+Au interaction.  However, the experimental
data ~\cite{stare,phenixe} at RHIC energy ($\sqrt{s_{NN}}$=200 GeV)
shows substantial suppression ($R_{AA}<1$) for $p_T\geq 2$ GeV indicating
substantial interaction of the plasma particles with charm and bottom quarks
from which electrons are originated through the process:
$c(b)$ (hadronization)${\longrightarrow}$ $D(B)$(decay)$\longrightarrow$
$e+X$. The loss of energy of high momentum heavy quarks propagating through
the medium created in Au+Au collisions causes a depletion of high $p_T$
electrons.

\begin{figure}[h]
\begin{center}
\includegraphics[scale=0.43]{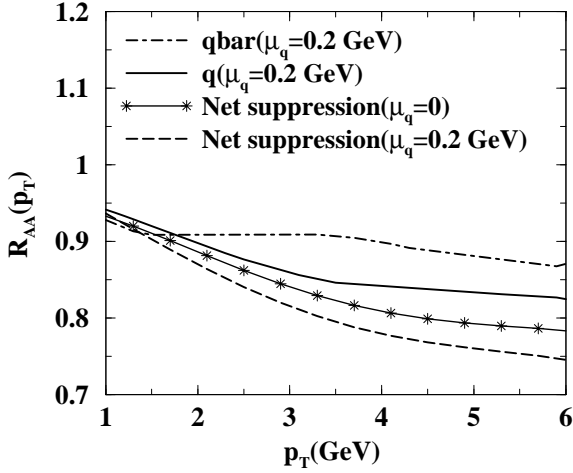}
\caption{The nuclear suppression factor $R_{AA}$ as a function
of $p_T$ due to the interaction of the charm quark (solid line)
and anti-quark (dashed-dot line) for $\mu_q=200$ MeV.
The net suppressions including the interaction of quarks, anti-quarks and
gluons for $\mu_q=200$ MeV (dashed line) and $\mu_q=0$ (with asterisk) 
are also shown.}
\label{fig5}
\end{center}
\end{figure}

\section{The initial conditions for the space-time evolution}
The nuclear suppression for heavy quarks depend on the parameters like initial
temperature  ($T_i$), thermalization time ($\tau_i$), equation of state (EOS)
and the transition temperature ($T_c$). 

We assume that the system reaches equilibration at a time
$\tau_i$ after the collision at temperature
$T_i$ which are related to the produced hadronic (predominantly mesons)
multiplicity 
through the following relation:
\be
T_i^{3}\tau_i \approx \frac{2\pi^4}{45\zeta(3)}\frac{1}{4a_{eff}}\frac{1}
{\pi R_A^2}\frac{dN}{dy}.
\label{eq6}
\ee
where
$R_A$ is the radius of the system,
$\zeta(3)$ is the Riemann zeta function and  $a_{eff}=\pi^2g_{eff}/90$ where
$g_{eff}$ ($=2\times 8+ 7\times 2\times 2\times 3\times N_F/8$) is the
degeneracy of quarks and gluons in QGP, $N_F$=number of flavours.

The value of the multiplicities for various $\sqrt{s_{\mathrm NN}}$
have been calculated from the Eq. below~\cite{kharzeev};
\be
\frac{dN}{dy}=\frac{dn_{pp}}{dy}\left[(1-x)\frac{<N_{part}>}{×2}+x<N_{coll}>
\right]
\ee
$N_{coll}$ is the number of
collisions and contribute  $x$ fraction to the multiplicity $dn_{pp}/dy$
measured in $pp$ collision. The number of participants,
$N_{part}$ contributes a 
fraction $(1-x)$ to $dn_{pp}/dy$, which is given by 
\be
\frac{dn_{pp}}{dy}=2.5-0.25ln(s)+0.023ln^2(s)
\ee
The values of $N_{part}$
and $N_{coll}$  are estimated for $(0-5\%)$ centralities by using Glauber Model.
The value of $x$ depends very weakly on $\sqrt{s_{\mathrm NN}}$~\cite{bbback},
in the present work we have taken $x=0.1$ for all the energies.

\begin{table}[h]
\caption{The values center of mass energy , $dN/dy$, initial temperature ($T_i$)
and quark chemical potential  - used
in the present calculations.}
\begin{tabular}{lccr}
\tableline
$\sqrt(s_{\mathrm NN})$(GeV)&$\frac{dN}{dy}$ &$T_i$(MeV) &$\mu_q$(MeV)\\
\tableline
39&617 &240  &62\\
27&592 &199  &70\\
17.3&574 &198  &100\\
7.7 &561 &197  &165\\
\tableline
\end{tabular} 
\end{table}

The time evolution of the temperature and the baryon density ($n_B$) have
been obtained by solving the following equations:
\be
\partial_\mu T^{\mu\nu}=0,\,\,\,\,\, \partial_\mu n_B^\mu=0
\ee
in (1+1) dimension with boost invariance along the longitudinal
direction~\cite{bjorken}. 
In the above equation $T^{\mu\nu}=(\epsilon+P)u^\mu u^\nu-g^{\mu\nu}P$, 
is the energy momentum tensor and
$n_B^\mu=n_Bu^\mu$ is the baryonic flux,
where $\epsilon$ is the energy density, $P$ is the pressure,
and $u^\mu$ is the hydrodynamic
four velocity. The radial co-ordinate dependence of $T$ and
$n_B$ have been parametrized as in Ref.~\cite{turbide}.  
The velocity of sound for the QGP phase is taken as $c_s=1/\sqrt{4}$.
Some comments on the effects of the radial flow are in order here. 
The radial expansion will increase the size of the system and hence
decrease the density of the medium. Therefore, with radial flow 
the heavy quark will traverse a larger path length in a  medium of 
reduced density. These two oppositely competing effects may have 
negligible effects on the nuclear suppression (see also~\cite{turbide}).  
Moreover, at lower collision energies (as in the present case) the
amount of radial flow will not be as substantial as in $\sqrt{s_{\mathrm NN}}
=200$ GeV.

The total amount of energy dissipated by a  heavy quark in the QGP
depends on the path length it traverses.
Each parton traverses different path length
which depends on the  geometry of the system and on the point 
where its is produced.
The probability that a parton is created at a point $(r,\phi)$
in the plasma depends on the number of binary collisions 
at that point which can be taken as~\cite{turbide}:
\be
P(r,\phi)=\frac{2}{\pi R^2}(1-\frac{r^2}{R^2})\theta(R-r)
\ee
where $R$ is the nuclear radius. 
A parton created at $(r,\phi)$ in the transverse plane
propagate a distance $L=\sqrt{R^2-r^2sin^2\phi}-rcos\phi$
in the medium. In the present work we use the following
equation for the averaging of the drag coefficient: 
\be
\Gamma=\int rdr d\phi P(r,\phi) \int^{L/v}d\tau\gamma(\tau)
\label{cgama}
\ee
where $v$ is the velocity of the propagating partons. 
Similar averaging has been performed  for the diffusion co-efficient.
For a static system the $T$ and $\mu_q$  dependence of the drag and
diffusion co-efficients of the heavy quarks enter via the
thermal distributions of light quarks, anti-quarks and gluons through
which it is propagating. However, in the present scenario
the variation of the temperature and the baryon density with time are 
governed by
the equation of state or the velocity of sound
of the thermalized system undergoing hydrodynamic
expansion. In such a scenario the quantities like $\Gamma$ (Eq.~\ref{cgama})
and hence $R_{\mathrm AA}$ becomes sensitive to velocity of sound
in the medium.

\begin{figure}[h]
\begin{center}
\includegraphics[scale=0.43]{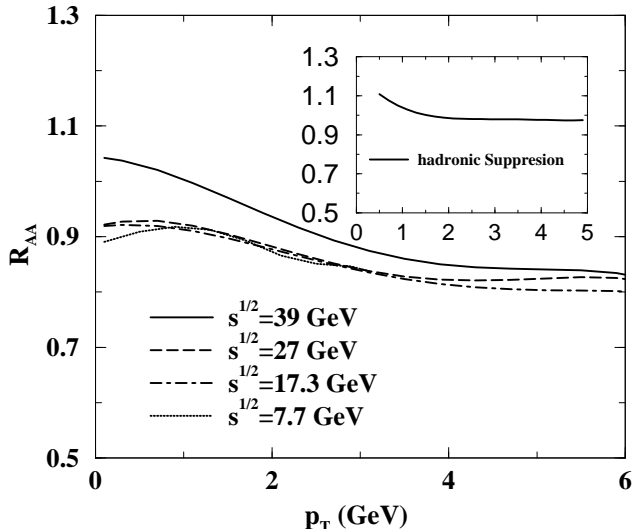}
\caption{Nuclear suppression factor, $R_{\mathrm AA}$ as function of $p_T$
for various $\sqrt{s_{NN}}$. Inset: the nuclear suppression factor
due to the interaction of $D$ meson in a thermal medium of pions and nucleons.
}
\label{fig6}
\end{center}
\end{figure}

\section{The nuclear suppression}
To demonstrate the effect of non-zero baryonic chemical
potential we evaluate $R_{\mathrm AA}$ for $\mu_q=200$ MeV
and $\mu_q=0$ for a given $T_i=200$ MeV. The
results are displayed in Fig.~\ref{fig5}
- representing the combined effects of 
temperature and baryon density on the viscous drag and diffusion. 
The viscous drag on the heavy quarks due to its interaction with
quarks is larger than that of its interactions with the anti-quarks
(Fig.\ref{fig1}).
Resulting in larger suppression in the former case than the later.
The net suppression of the electron spectra from the Au+Au collisions 
compared to p+p collisions is effected by quarks, anti-quarks and gluons.
The results for net suppressions are displayed for $\mu_q=200$ MeV (dashed
line) and $\mu_q=0$ (with asterisk). 
The experimental detection of the
non-zero baryonic effects will shed light on the net baryon density
(and hence baryon stopping) in the central rapidity region.
However, whether the effects of non-zero baryonic 
chemical potential is detectable or not will depend on the 
overall experimental performance.

The results for $R_{AA}$ are shown in Fig.~\ref{fig6}
for various $\sqrt{s_{NN}}$ with inputs from table I.
We observe that 
at large $p_T$ the suppression is similar for all energies
under consideration. This is because the collisions
at high $\sqrt{s_{NN}}$  are associated with large temperature but
small baryon density at mid-rapidity- which is compensated by  large 
baryon density and small  temperature at low $\sqrt{s_{NN}}$ collisions.  
Low $p_T$ particles predominantly originate from low temperature and 
low density part
of the evolution where drag is less and so is the nuclear suppression.

So far we have discussed the suppression of the non-photonic electron
produced in nuclear collisions due to the propagation of the heavy quark
in the the partonic medium  in the pre-hadronization era. However, 
the suppression of the $D$ mesons in the post hadronization era 
(when both the temperature and density are lower than the partonic
phase) should
in principle be also taken into account.  We have estimated $R_{AA}$
for $D$ mesons due to ts interaction with pions~\cite{dpi}
and nucleons~\cite{dnucleon} and found that it has a value
value closer to unity, indicating the fact that the hadronic
medium (of pions and nucleons)  is unable to drag the $D$ mesons.
Therefore, the measured depletion in $R_{AA}$ for the non-photonic electron
will indicate the presence of partonic medium and the amount of depletion
may the used to characterize the thermal medium.

\begin{figure}[h]
\begin{center}
\includegraphics[scale=0.43]{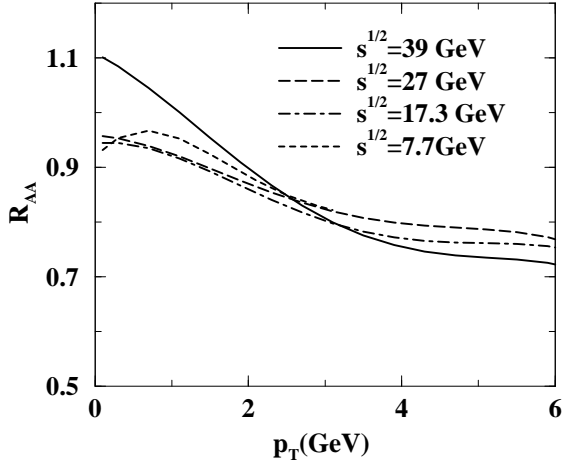}
\caption{Same as Fig.~\protect{\ref{fig6}} with enhancement of cross section 
by a factor of 2.}
\label{fig7}
\end{center}
\end{figure}

It has been shown in~\cite{molnar} that
a large enhancement of the pQCD cross section is required for
the reproduction of experimental data on elliptic flow at RHIC
energies.
In our earlier work~\cite{Das} we have evaluated the 
$R_{\mathrm AA}$ for non-photonic single electron spectra 
resulting from the semileptonic decays of hadrons containing 
heavy flavours and observed that the data from RHIC collisions
at $\sqrt{s_{NN}}=200$ GeV are well reproduced by enhancing the
pQCD cross sections by a factor 2  and with an equation of state
$P=\epsilon/4$. In the same spirit we evaluate $R_{\mathrm AA}$ 
with twice enhanced pQCD cross section
and  keeping all other quantities unaltered~(Fig.~\ref{fig7}).
The results in Fig.~\ref{fig7} show stronger  suppression
as compared to the results displayed in Fig.~\ref{fig6}, but 
it is similar in all the energies under consideration.

\section{Summary and conclusions}
We have studied the effects of baryonic chemical potential and temperature on
the drag and diffusion coefficients of heavy quarks moving in a thermalized
system of quarks and gluons. We have observed that both the drag and diffusion
coefficients increase with temperature and chemical potential.
When  we have enhanced the pQCD 
cross section for the interaction of the heavy quarks with the thermal system 
by a factor of two - the resulting suppressions in $R_{AA}$  are between 
$20\% - 30\%$ for $\sqrt{s_{NN}}=39-7.7$ GeV. 
The  radiative energy loss~\cite{pqm,glv,zoww,bdps,salgado}
(see ~\cite{revieweloss} for a review) of heavy quarks is
suppressed due to dead cone effects and has
been neglected in the present work. Moreover, at low collisions energies
the collisional loss~\cite{hvh,ko,gossiaux}
is dominant over its radiative counter part (see~\cite{akdm} for details).
It may be mentioned here that the 
theoretical formalism, the FP equation is applicable better 
for heavy quarks than light quarks and gluons (because of
their frequent productions and annihilations).
However, the production of charm
and bottom quarks are smaller at low energy collisions 
making the measurements of non-photonic 
single electron spectra and hence $R_{\mathrm AA}$  
for heavy quarks difficult.
The detection of the non-zero baryonic chemical potential effects observed 
in the present work through the nuclear suppression factor 
will help in determining the net baryon density
(and hence baryon stopping) in the mid-rapidity region.
However, whether such effects  
is detectable experimentally or not will depend on the 
overall experimental performance.


{\bf Acknowledgment:}
We thank Bedangadas Mohanty and Jajati K Nayak for useful discussions.
This work is supported by DAE-BRNS project Sanction No. 2005/21/5-BRNS/2455.


\begin{thebibliography}{99}
\bibitem{lowephenix} T. Sakaguchi (PHENIX collaboration), 
arXiv:0908.3655 [hep-ex].

\bibitem{lowestar} B. I. Abelev (Star Collaboration),
arXiv:0909.4131 [nucl-ex]

\bibitem{cbm} J. M. Heuser (CBM collaboration), J. Phys. G:
Nucl. Phys. {\bf 35}, 044049 (2008). 



\bibitem{moore} G. D. Moore and D. Teaney, Phys. Rev. C {\bf 71}, 064904
 (2005).

\bibitem{baier} R. Baier, A. H. Mueller, D. Schiff and D. T. Son,
Phys. Lett. B {\bf 539}, 46  (2002).

\bibitem{japrl} J. Alam, S. Raha and B. Sinha, Phys. Rev. Lett. {\bf 73}, 1895
(1994).

\bibitem{shuryak} E. Shuryak, Phys. Rev. Lett. {\bf 68}, 3270 (1992).

\bibitem{landau} E. M. Lifshitz and L. P. Pitaevskii, Physical Kinetics,
Butterworth-Hienemann, Oxford 1981.

\bibitem{balescu} R. Balescu, Equilibrium and Non-Equilibrium Statistical
Mechanics (Wiley, New York, 1975).


\bibitem{sc} S. Chakraborty and D. Syam, Lett. Nuovo Cim. {\bf 41}, 381 (1984).

\bibitem{svetitsky} B. Svetitsky, Phys. Rev. D {\bf 37}, 2484( 1988).

\bibitem{rapp} H. van Hees, R. Rapp, Phys. Rev. C,{\bf 71}, 034907 (2005).

\bibitem{turbide} S. Turbide, C. Gale, S. Jeon and G. D. Moore,
Phys. Rev. C {\bf 72}, 014906 (2005).

\bibitem{bjoraker} J. Bjoraker and R. Venugopalan, Phys. Rev. C {\bf 63},
024609 (2001).

\bibitem{npa1997} P. Roy, J. Alam, S. Sarkar, B. Sinha and S. Raha,
Nucl. Phys. A {\bf 624}, 687 (1997).

\bibitem{munshi} M  G. Mustafa and  M. H. Thoma, Acta Phys. Hung. A
{\bf 22}, 93 (2005).

\bibitem{rma} P. Roy, A. K. Dutt-Mazumder and J. Alam, Phys. Rev. C {\bf 73},
044911 (2006).

\bibitem{Das}S. K Das, J. Alam and P. Mohanty, 
Phys. Rev. C {\bf 80}, 054916 (2009).

\bibitem{ristea} O. Ristea (for the BRAHMS collaboration)
Romanian Reports in Physics, {\bf 56}, 659(2004)

\bibitem{andronic} A. Andronic, P. Braun-Munzinger and J. Stachel, 
Nucl. Phys. A {\bf 772}, 167 (2006).  


\bibitem{gossiaux}  P. B. Gossiaux and J. Aichelin, Phys. Rev. C
{\bf 78}, 014904 (2008).

\bibitem{rapphees} H. van Hees, M. Mannarelli, V. Greco and R. Rapp,
Phys. Rev . Lett. {\bf 100}, 192301 (2008).

\bibitem{caron-huot} S. Caron-Huot and G. D. Moore, Jour. High Ener. Phys.
{\bf 0802}, 081 (2008);
S. Caron-Huot and G. D. Moore, Phys. Rev. Lett.
{\bf 100}, 052301 (2008).

\bibitem{matteo} M. Cacciari, P. Nason and R. Vogt, Phys. Rev. Lett.
{\bf 95}, 122001 (2005).

\bibitem{pqcd} R. D. Field, Application of Perturbative QCD, Addison-Wesley
Pub. Company, N.Y. 1989.

\bibitem{combridge} B.L. Combridge, Nucl.Phys.B {\bf 151}, 429 (1979).

\bibitem{peterson} C. Peterson {\it et al.},  Phys. Rev. D {\bf 27}, 105 (1983).

\bibitem{gronau} M. Gronau, C. H. Llewellyn Smith,
T. F. Walsh, S. Wolfram and T. C. Yang,
Nucl. Phys. B {\bf 123}, 47 (1977).

\bibitem{ali} A. Ali, Z. Phys. C {\bf 1}, 25 (1979).

\bibitem{akc} A. Chaudhuri, nucl-th/0509046.

\bibitem{stare} B. I. Abeleb {\it et al.} (STAR Collaboration), Phys. Rev.
Lett. {\bf 98}, 192301 (2007).

\bibitem{phenixe} S. S. Adler {\it et al.} (PHENIX Collaboration),
Phys. Rev. Lett. {\bf 96}, 032301 (2006).

\bibitem{kharzeev} D. Kharzeev and M.Nardi, Phys. Lett. B, {\bf 507}, 
121 (2001).

\bibitem{bbback} B. B. Back {\it et al.} (PHOBOS Collaboration),
Phys. Rev. C {\bf 70}, 021902 (2004).

\bibitem{bjorken} J. D. Bjorken, Phys. Rev. D {\bf 27}, 140 (1983).

\bibitem{dpi} C. Fuchs, B. V. Martemyanov, A. Faessler and
M. I. Krivoruchenko, Phys. Rev. C {\bf 73}, 035204 (2006).

\bibitem{dnucleon} J. Haidenbauer, G. Krein, U. -G. Meissner and
A. Sibirtsev, Eur. Phys. J. A {\bf 33}, 107 (2007). 

\bibitem{molnar} D. Molnar and M. Gyulassy, Nucl. Phys. A {\bf 697}, 495 (2002).

\bibitem{pqm} A. Dainese, C. Loizides and G. Paic, Eur. Phys. J. C {\bf 38}
(2005) 461; C. Loizdes, {\it ibid.} {\bf 49}, 339 (2007).

\bibitem{glv} M. Gyulassy, P. Levai and I. Vitev, Nucl. Phys.
B {\bf 571}, 197 (2000);
M. Gyulassy, P. Levai and I. Vitev, Phys. Rev. Lett,
{\bf 85}, 5535 (2000);
M. Gyulassy and X.-N. Wang, Nucl. Phys. B {\bf 420}, 583 (1994).

\bibitem{zoww} H. Zhang, J. F. Owens, E. Wang and X. N. Wang, Phys.
Rev. Lett. {\bf 98}, 212301 (2007).

\bibitem{bdps} R. Baier, Y. L. Dokshitzer, S. Peigne and D. Schiff,
Phys. Lett. B {\bf 345}, 277 (1995);
R. Baier, Y. L. Dokshitzer, A. H. Mueller, S. Peigne and
D. Schiff, Nucl. Phys. B {\bf 531},  403 (1998);

\bibitem{salgado} C. A. Salgado and U. A. Wiedemann,
Phys. Rev. Lett. {\bf 89}, 092303 (2002).

\bibitem{revieweloss} P. Jacobs and  X. N. Wang,
Prog. Part. Nucl. Phys. {\bf 54}, 443 (2005);
R. Baier, D. Schiff, B. G. Zakharov,
Ann. Rev. Nucl. Part. Sci. {\bf 50}, 37 (2000).

\bibitem{hvh} H. van Hees, M. Mannarelli, V. Greco and R. Rapp,
Phys. Rev. Lett. {\bf 100}, 192301 (2008).

\bibitem{ko} C. M. Ko and W. Liu, Nucl. Phys. A {\bf 783}, 23c (2007).

\bibitem{akdm} A. K. Dutt-Mazumder, J. Alam, P. Roy and B. Sinha,
Phys. Rev. D {\bf 71}, 094016 (2005).  

\end{thebibliography}
\end{document}